# Imaging Electrons in a Magnetic Field

Katherine E. Aidala[a*], Robert E. Parrott[b,+], E. J. Heller[b,c], R. M. Westervelt[a,b]

[a] *Division of Engineering and Applied Sciences, Harvard University, Cambridge, MA, 02138*

[b] *Department of Physics, Harvard University, Cambridge, MA, 02138*

[c] *Department of Chemistry and Chemical Biology, Harvard University, Cambridge, MA 02138*

**Abstract**

We present simulations of an imaging mechanism that reveals the trajectories of electrons in a two-dimensional electron gas (2DEG), as well as simulations of the electron flow in zero and small magnetic fields. The end goal of this work is to implement the proposed mechanism to image the flow of electrons inside a ballistic electron device from one specific point (A) to another (B) in a 2DEG, using a low temperature scanning probe microscope with a charged tip. The tip changes the electron density in the 2DEG beneath it and deflects the electrons traveling nearby, thereby changing the conductance from point A to point B. The simulations presented here show that by measuring the transmission of electrons from A to B versus tip position, one can image the electron flow. This forward scattering mechanism is well suited for imaging in a magnetic field, in contrast to previous probes that depended on backscattering. One could use this technique to image cyclotron orbits in an electron focusing geometry, in which electrons travel from point A to point B in semi-circular paths bouncing along a wall. Imaging the motion of electrons in magnetic fields is useful for the development of devices for spintronics and quantum information processing.

## 1. Introduction

Scanning probe microscopy (SPM) has been successful in locally probing two-dimensional electron systems [1]. Previous imaging experiments investigated quantum dots [2,3,4] and quantum point contacts (QPCs) [5,6,7]. Earlier work performed by Westervelt and coworkers [6] on a QPC in a two-dimensional electron gas (2DEG) produced high resolution images of electron flow from a QPC by depleting the electron density below the SPM tip to create a divot that backscattered electrons from the QPC (point A) back to the QPC (A). It would be very useful to be able to image the flow of electrons from point A to point B and in a magnetic field for the design of new devices. Understanding electron motion in a magnetic field will be important for spintronic devices and quantum information processing.

---

[*] Corresponding author. Tel.: 617-495-9598; e-mail: aidala@fas.harvard.edu

[+] Generated all of the simulations in this paper

One can image electron flow from point A to point B in a magnetic field by using a SPM tip to reduce the density of the electrons beneath the tip. In the single electron picture, this creates an effective local potential hump in the 2DEG, which acts as a classical diverging lens. Those electrons that pass nearby will be deflected by this perturbation, thereby changing the conductance between the two points. The strength of the perturbation created by the tip will determine the extent to which electrons are deflected. One can obtain an image of the electron flow by measuring this change in the conductance while scanning the tip over the sample. This is demonstrated in a series of classical simulations, first in zero magnetic field between two QPCs that face each other, similar to the experiment performed in [7]. We then simulate magnetic focusing of the electron flow between two QPCs facing in the same direction, and obtain images of the electron paths in the magnetic field.

## 2. Results

Figure 1 demonstrates that a SPM tip that slightly depletes the 2DEG beneath it can be used to image electrons moving from one QPC to another. The white horizontal lines at the top and bottom of the image are the location of the gates forming the QPCs. This geometry was first used to explore electron collimation and modal structure in a QPC [8, 9]. Figures 1a, b and c display a classical simulation of electron paths. A sheet of trajectories is launched from the bottom QPC and is propagated across the area just as one uses ray-tracing in optics. Figure 1a shows a smooth distribution of electrons, forming a wedge emanating from the QPC. These trajectories subtend an angle of 30 degrees. The 100 nm wide circle is the location of the tip perturbation, which is approximated by a Gaussian with a full-width at half-max set here to be 20% of the Fermi energy. Electrons near the tip are deflected away, and this defocusing results in a shadow behind the tip, bordered by caustics. The tip acts as a transmitting lens with no absorption, so that if there is a decrease in the classical density in one location (the shadow) there must be a corresponding increase elsewhere (the caustics). The caustic that reaches the bottom right corner of the figure has bounced off the top gate. The solid line in figure 1d plots the transmission of electrons from the bottom QPC to the top with respect to the location of the tip, which is placed at y = 1000 nm and scanned along x. The peaks in the plot are where either caustic bounding the shadowed region pass by the top QPC. The center dip is the shadow. The transmission is defined as the fraction of electrons that make it through the top QPC. The transmission T provides a classical approximation to the conductance of the system, such that $G = T*2e^2/h$ if the bottom QPC is on the first plateau.

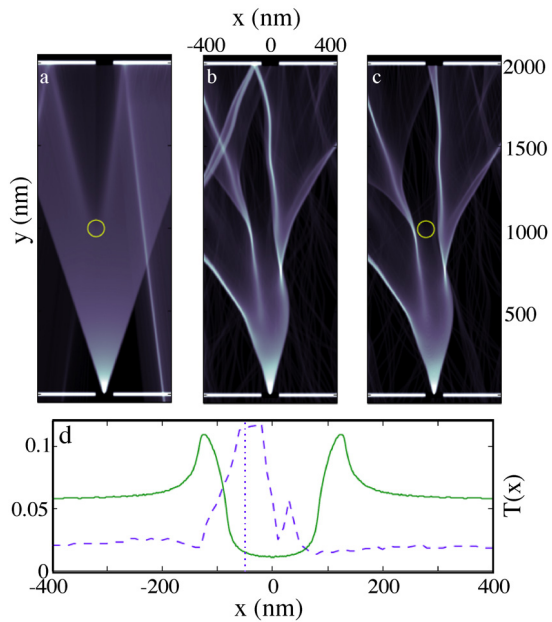

**Figure 1** A SPM tip can image electron flow between two QPCs. 1a – c are ray-tracing simulations of electron flow. 1a has no disorder, and the white circle indicates the position of a tip that decreases the density of the 2DEG beneath it. 1b includes disorder, causing branches to form. 1c places the tip perturbation in the disorder of 1b, shifting a branch so that it is transmitted. 1d plots the transmission versus location of the tip along the x-axis, at y=1000 nm. The symmetric solid line is from 1a, the dotted line is from 1b. The dotted line reveals the location of the branches.

Figure 1b adds random disorder to the picture, without the presence of the tip. Small angle scattering results in visible branches, as described in [6]. Note the branch that hits the top QPC just to the left of the opening, and is reflected. Though it is not initially transmitted through the top QPC, when the tip is placed at the location in Figure 1c, it will be imaged because the tip deflects the branch *towards* the top QPC. Electrons passing near the tip can be deflected both into and away from the collector, and any electrons that are transmitted through the second QPC for some location of the tip can be imaged. The dashed line in Fig1d demonstrates this. Note that the peak in transmission occurs when the tip is located at the branch, indicated by the dotted vertical line.

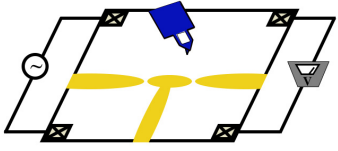

**Figure 2** Schematic of electron focusing measurement. Electrons are sourced from one QPC and the voltage resulting from injected electrons is measured across the second. A SPM can be used to image trajectories between the two.

An interesting geometry to image in a magnetic field is the traditional focusing geometry [10]. Two QPCs are side by side separated by a narrow distance smaller than the mean free path of the electrons, as shown in Figure 2. Spin polarized currents have been reported due to spin-orbit effects in similar geometries [11,12]. One sources a current through the first QPC that acts as an emitter of electrons, and measures the voltage across the second that acts as a collector. This voltage indicates the total transmission of the system because it drives the current that balances the flow injected into the second QPC. The ratio of the current into the second QPC to the current out of the first QPC gives the transmission and total conductance of the system.

Upon turning on a magnetic field perpendicular to the plane of the electrons, the emitted electrons will follow a circular path determined by the cyclotron radius,

$$r_c = \frac{mv}{eB} \qquad (1)$$

where m is the effective mass, and v the velocity. Though the electrons passing through the first point contact are diverging, the magnetic field focuses the electron trajectories. The middle gate interrupts the paths of the electrons at a point of near convergence, and the electrons travel in semicircular paths bouncing along the wall. When one of these bounces intersects the collector QPC, a peak will be seen in the measured transmission. This occurs when

$$B = \frac{2mv}{eL} \qquad (2)$$

where L is the spacing between the two QPCs.

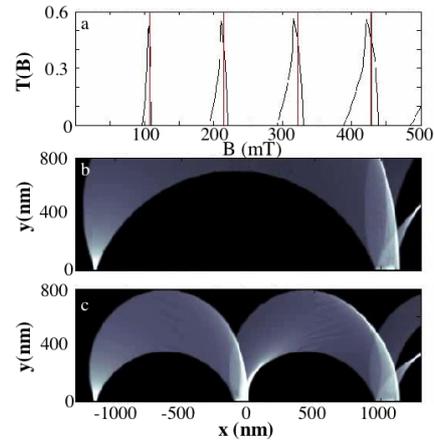

**Figure 3** 3a plots the transmission from one QPC to the second versus magnetic field in the focusing geometry of Figure 2. The vertical lines indicate the expected location of the peaks from equation (2). 3b displays a ray-tracing simulation of electron flow at the first peak, 3c corresponds to the second peak.

The expected peaks are calculated in Figure 3a by using ray-tracing techniques and computing the fraction of trajectories transmitted between the QPCs in a magnetic field, without disorder. The vertical lines are positioned where the distance between point contacts is equal to a multiple of the cyclotron diameter. The corresponding electron flow at these first two points is displayed in 3b and 3c. The peak in transmission is slightly offset from the ideal value, and the bases of subsequent peaks broaden as expected [10]. The shape of the peaks is determined by the collimation and the imperfect convergence of the trajectories. Each bounce adds to the imperfect convergence and the spread in the trajectories increases. This is seen in figure 3c where the second bounce covers a distinctly wider distance than the first.

The focusing geometry can be imaged with a charged SPM tip, as demonstrated in Figure 4. Figure 4a is identical to Figure 3c with the addition of the diverging tip perturbation, as described previously, revealing the shadow behind the tip. After the bounce, the caustics converge to form a focal point, indicated by the white arrow, past which the shadow appears once again. The features created by the tip are essentially replicated for each semicircular segment. Figure 4b includes a *converging* tip potential, and therefore bends nearby electron trajectories towards one another. This is equivalent to a positively charged tip increasing the density below. Instead of the shadow, there is a bright region where trajectories converge. Past this focal point, these trajectories once again diverge, and create the shadow region after the bounce. The arrows indicate the focal point, and its reflection after the bounce. Figures 4a and 4b are opposites of one another, the charge on the tip determining on which side of the tip the focus lies. Scanning the tip over the entire area images the semicircular paths. Figure 4c is a classical simulation of such a SPM scan. The transmission is calculated and plotted at each location of the tip. While the exact contrast would be dependent on the size of the perturbation, the cyclotron orbits are easily visualized. Disorder would complicate this picture as some branches would be deflected out of transmission, and others into transmission, such that the contrast varies and some branches will be more readily visualized than others. There may be values of magnetic fields and sets of disorder potentials for which some branches cannot be seen. For magnetic field values close to the focusing peaks, nearly all the electrons are initially transmitted and can be imaged by scanning the SPM.

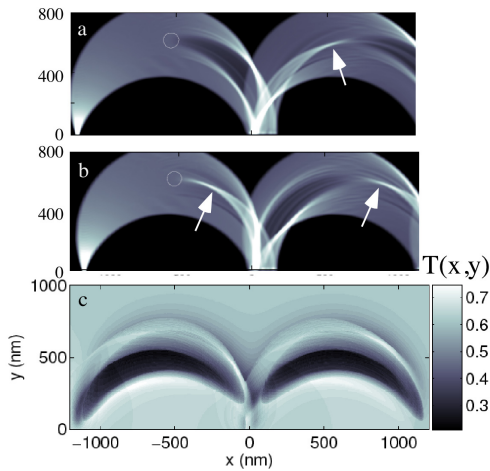

**Figure 4** 4a shows the change in electron trajectories in the presence of a positively charged tip that slightly increases the electron density beneath it. 4b shows the effect of a negatively charged tip that slightly decreased the electron density beneath it. The arrows indicate the focal points that result from the tip. 4c is a simulation of an entire tip scan. Each point is the fraction of electrons that are transmitted to the second QPC given the presence of the tip at that location.

## 3. Conclusions

Knowledge of how electrons move from one point to another in small electronic structures is important for creating future spintronic and quantum information processing devices. Here was have presented an analysis of a scanning probe microscope technique to image flow with or without a magnetic field in a 2DEG. By changing the electron density beneath the SPM slightly, one can successfully image electron flow between two quantum point contacts.

**Acknowledgemets:** The authors wish to thank Diego Vaz Bevilaqua and Michael Stopa for helpful discussions. This work was supported in part by ARO grant W911NF-04-1-0343.